\documentclass[sigconf,nonacm]{acmart}

\usepackage{todonotes}
\usepackage{multirow}
\usepackage{cleveref}
\usepackage{subcaption}
\usepackage{enumitem}

\renewcommand\footnotetextcopyrightpermission[1]{} % Optional: suppress copyright notice for submission

\AtBeginDocument{%
  }

\setcopyright{acmlicensed}
\copyrightyear{2018}
\acmYear{2018}
\acmDOI{XXXXXXX.XXXXXXX}
\acmConference[]{Make sure to enter the correct
  conference title from your rights confirmation email}{}{}

\acmISBN{978-1-4503-XXXX-X/2018/06}

\usepackage{xspace}
\usepackage[utf8]{inputenc}
\usepackage{tipa}
\newcommand{\system}{\textsc{Sa\textsubdot{m}sāra}\xspace} %here enter the proposed system name
\begin{document}

\title{[Vision Paper] Towards a Multimodal Stream Processing System }

\author{Uélison Jean Lopes dos Santos}

\affiliation{%
  \institution{TU Darmstadt and DFKI}
  %\city{Germany}
  %\state{Darmstadt}
  %\country{Germany}
  \country{}
}
%\email{uelison.santos@dfki.de}

\author{Alessandro Ferri}
\affiliation{%
  \institution{TU Darmstadt}
  %\city{Germany}
  %\state{Darmstadt}
  %\country{Germany}
  \country{}
}

\author{Szilard Nistor}
\affiliation{%
  \institution{TU Darmstadt and DFKI}
  %\city{Germany}
  %\state{Darmstadt}
  %\country{Germany}
  \country{}
}

\author{Riccardo Tommasini} %add Riccardo
\affiliation{%
  \institution{INSA Lyon}
  %\city{Germany}
  %\state{Darmstadt}
  %\country{France}
  \country{}
}

\author{Carsten Binnig}
\affiliation{%
  \institution{TU Darmstadt and DFKI}
  %\city{Germany}
  %\state{Darmstadt}
  %\country{Germany}
  \country{}
}

\author{Manisha Luthra}
\affiliation{%
  \institution{TU Darmstadt and DFKI}
  %\city{Germany}
  %\state{Darmstadt}
  %\country{Germany}
  \country{}
}

\begin{abstract}
In this paper, we present a vision for a new generation of multimodal streaming systems that embed MLLMs as first-class operators, enabling real-time query processing across multiple modalities. Achieving this is non-trivial: while recent work has integrated MLLMs into databases for multimodal queries, streaming systems require fundamentally different approaches due to their strict latency and throughput requirements. Our approach proposes novel optimizations at all levels, including logical, physical, and semantic query transformations that reduce model load to improve throughput while preserving accuracy. We demonstrate this with \system{}, a prototype leveraging such optimizations to improve performance by more than an order of magnitude. Moreover, we discuss a research roadmap that outlines open research challenges for building a scalable and efficient multimodal stream processing systems.
\end{abstract}

\maketitle

\vspace{-3ex}
\section{Introduction}
\label{sec:intro}

\begin{figure*}[ht]
  \centering
  
  \begin{subfigure}[b]{0.732\textwidth}
    \includegraphics[width=0.5\textwidth]{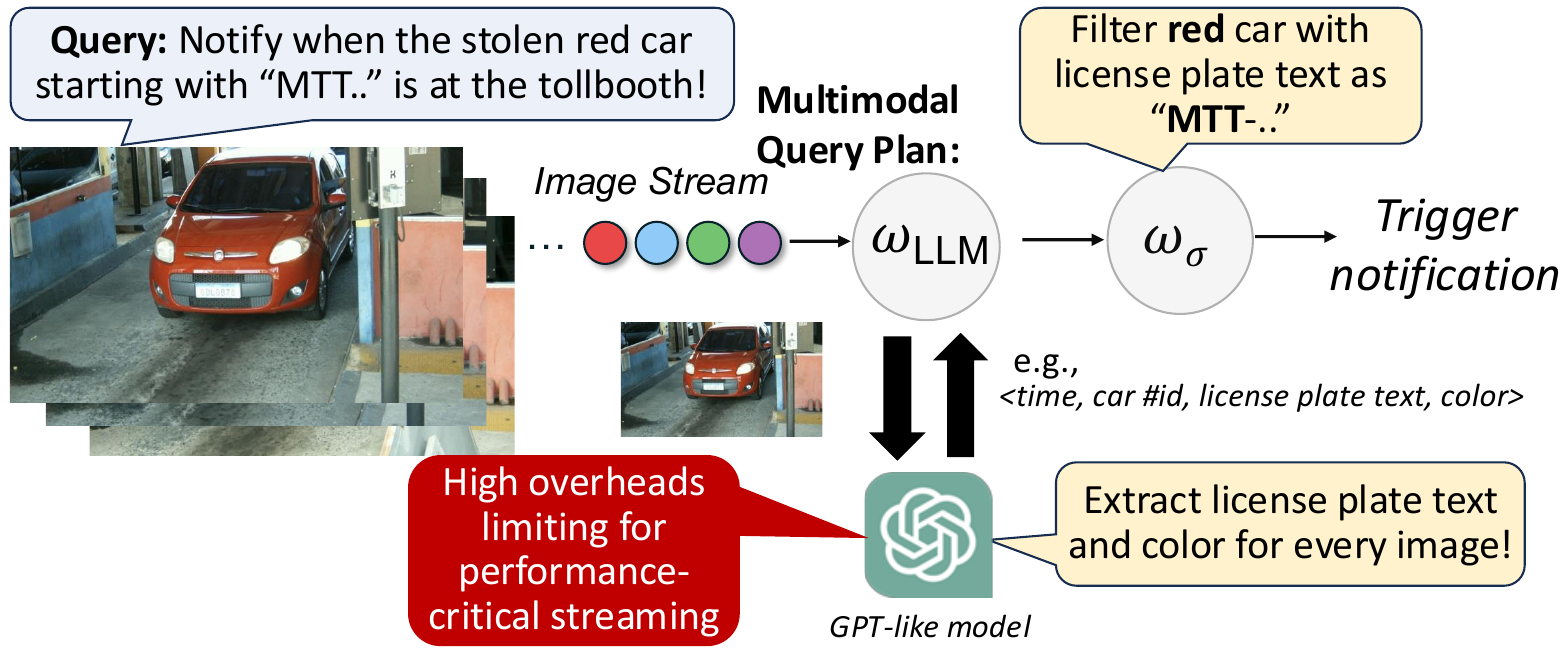}
    \includegraphics[width=0.5\textwidth]{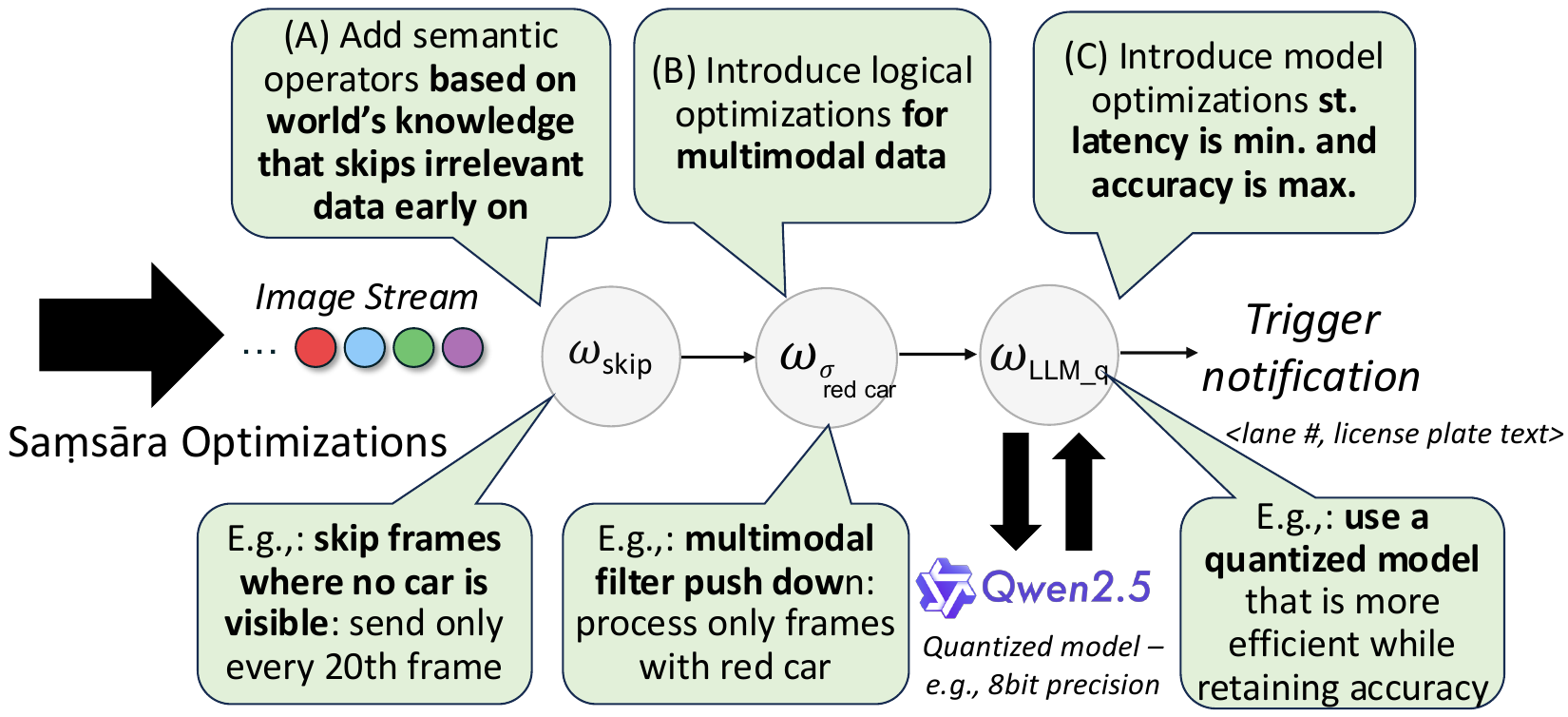} 
    \caption{Example query before and after \system\ optimization.}
    \label{fig:a}
  \end{subfigure}
  \hspace{1ex}
  \begin{subfigure}[b]{0.25\textwidth}
    \includegraphics[width=\textwidth]{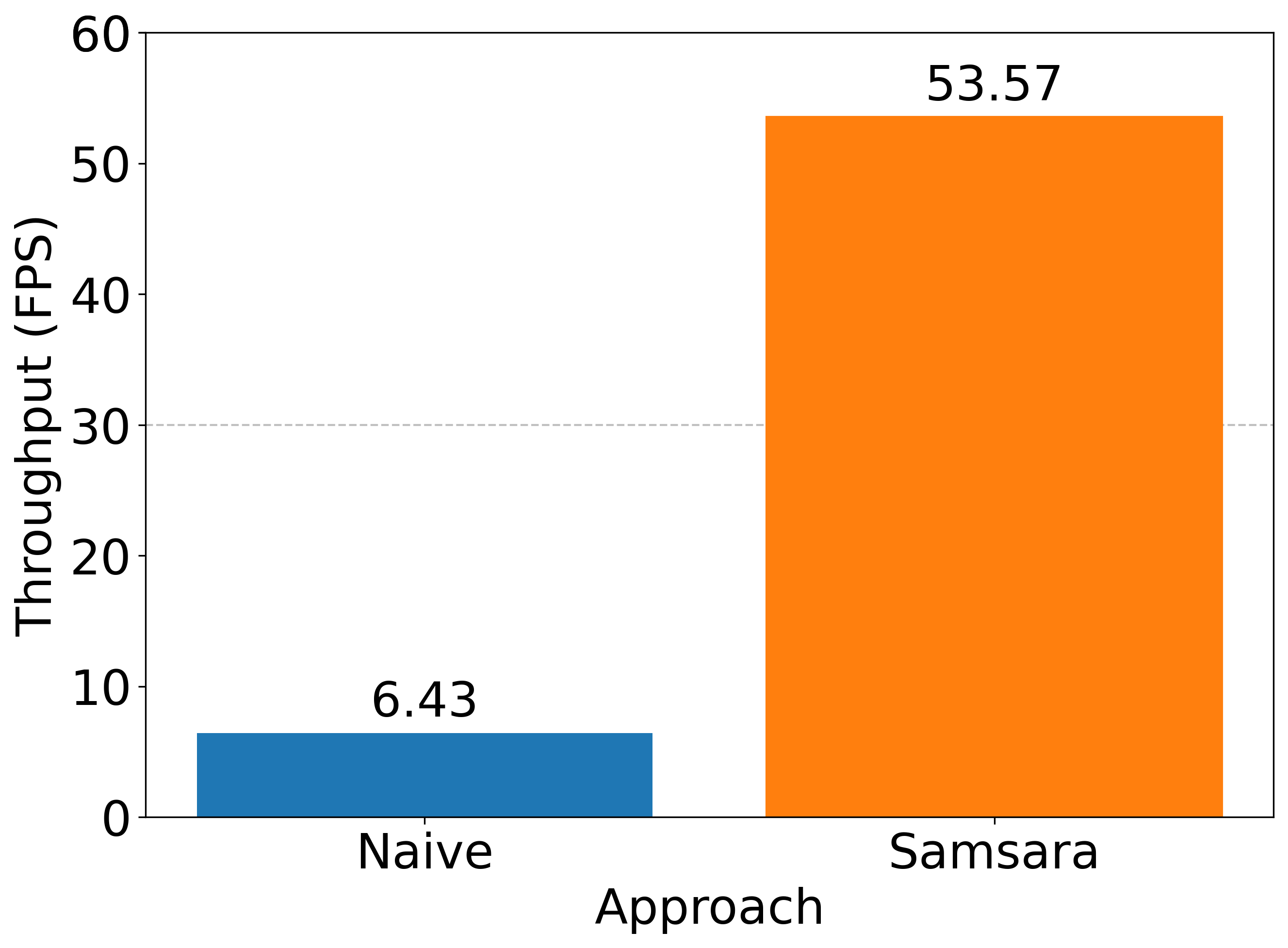}
    \caption{Naive vs. optimized execution. }
    \label{fig:b}
  \end{subfigure}
  \vspace{-5ex}
  \caption{Illustrating multimodal plan optimizations proposed by \system\ in (a) with evaluation results in (b). Naively integrating MLLMs into stream processing leads to extremely low performance. With \system\ and its novel optimizations for multimodal data streams, we achieve up to $9\times$ performance improvements for this example.}
  \label{fig:latency_vs_accuracy}
  \vspace{-3ex}
\end{figure*}

\textbf{Success and Limitation of Streaming Systems.} Stream processing systems (SPSs) have become fundamental in numerous industries, providing real-time data processing capabilities that power applications in finance~\cite{Sadoghi2010finance}, entertainment~\cite{netflixFlink}, healthcare~\cite{Wang2010healthcare}, and the Internet-of-Things (IoT)~\cite{chaudhary2020}. These systems continuously ingest, process, and analyze data streams, enabling timely decision-making based on the most current information. 
%A distinguishing feature of stream processing is the need to evaluate \textit{continuous queries} over \textit{continuous data streams}, specifying what data to process and how to transform or analyze it in real time. 
%\textbf{Restrictions of Streaming Systems.} 
However, despite their success, existing SPSs pose significant limitations that restrict broader applications. In particular, streaming systems today are designed to handle structured data but are not equipped to process diverse, unstructured, or multimodal data types, such as image streams from cameras or other modalities, including audio sensors.

\textbf{Leveraging multimodal LLMs for Streaming.} Recent advances in multimodal large language models (MLLMs) have demonstrated remarkable abilities to process and integrate data across multiple modalities, such as images, text, and audio, providing contextual understanding that spans these diverse data types. 
As such, it seems an appealing idea to use MLLMs as a building block within stream processing systems.
In fact, integrating MLLMs into stream processing opens up new possibilities to natively support rich queries on modalities beyond structured data, enabling systems to process and interpret multimodal data in real-time. Such capabilities could significantly enhance streaming systems in applications ranging from traffic monitoring using camera streams and autonomous vehicles to sports analytics and robotics.

\textbf{Towards Multimodal Stream Processing.} We envision a new generation of SPSs that can seamlessly query across multiple modalities (video, audio, text) by embedding MLLMs as first-class operators in the query plan. Figure \ref{fig:a} illustrates an example query where a user aims to detect a stolen car at a toll station using a camera stream. While current data systems \cite{urban2023caesura,10.14778/3749646.3749685,liu2025palimpzest,google-ai,snowflake-ai} have already proposed integrating MLLMs for querying multimodal data, streaming systems pose fundamentally different challenges. Streaming environments require extremely low latency and high throughput, which makes naive MLLM integration, simply by calling out to the model, infeasible. Achieving acceptable performance requires carefully optimized query execution. Unlike batch-oriented databases, streaming systems cannot tolerate long inference times, making efficient multimodal stream processing a highly non-trivial problem.  

\textbf{Our Vision} In this paper, we present our vision of how MLLMs should be combined with streaming systems and use their capabilities to process modalities like images out-of-the-box, while satisfying high-throughput and low-latency demands.
To enable such efficient multimodal streaming, we introduce the vision of a \emph{super-optimizer}---a new class of optimizers designed specifically for multimodal stream processing. Unlike traditional query optimizers, a super-optimizer generates deeply optimized query plans tailored to one particular query and data stream. 
We argue that this super optimization pays off since streaming queries are long-running, and this upfront effort leads to high performance benefits while running the query.
For this, a super-optimizer adds several novel optimization steps. For example, we introduce a new phase in optimization called \emph{semantic optimization} in addition to logical and physical optimizations, which optimizes plans based on a semantic understanding of data and queries to specialize the plan for a given scenario.
For instance, in a traffic monitoring scenario, understanding that cars appear one after another allows a streaming system to skip redundant frames and avoid unnecessary, expensive inference. 
Moreover, a super-optimizer employs techniques such as aggressive model specialization and pruning to reduce inference load. 

\textbf{Gains and Challenges.} We demonstrate these ideas in our prototype \system{}: a super-optimizer, which we integrated into an existing streaming system Apache Flink~\cite{carbone2015flink}. In contrast to a naive evaluation in Flink without our optimizer, we can achieve orders-of-magnitude throughput improvements---in our example, from $6$ to $53$ images per second---showcasing the potential of super-optimizers for multimodal streaming.
However, building such a \emph{super-optimizer} to enable multimodal streaming systems that work generally for all kinds of data streams and queries is far \emph{from trivial and requires extensive research}.
In fact, building robust optimizers for structured data has taken decades, and we believe that this paper can only be a starting point for multimodal streaming optimization powering efficient multimodal stream processing on top of MLLMs.

\textbf{Outline.} The remainder of the paper is organized as follows. Section~\ref{sec:vision} presents our vision for multimodal stream processing, highlighting the intuition behind the novel optimizations. Section~\ref{sec:case_study} provides a case study that illustrates these optimizations through concrete examples and explains details for the individual optimization phases. Finally, Section~\ref{sec:road_ahead} outlines the road ahead by discussing open challenges and future directions.  
\vspace{-2ex}
\section{Vision: Multimodal Stream Processing}
\label{sec:vision}

Our vision is to enable a new generation of stream processing systems capable of understanding and querying across multiple modalities---text, images, and audio---by leveraging recent advances in multimodal large language models (MLLMs) as first-class operators within query plans. 
However, simply integrating MLLMs into query plans as done in batch-oriented systems such as \textsc{CAESURA}~\cite{urban2023caesura}, Lotus~\cite{10.14778/3749646.3749685}, or DocETL~\cite{shankar2024docetl} is infeasible in a streaming context which demands low latency and high throughput. 

\textbf{A Super-Optimizer for Multimodal Streaming.} 
To overcome these challenges, we propose \system{}, a novel \emph{super-optimizer} that aggressively transforms multimodal query plans into efficient, low-latency, and accurate execution plans. 
The core idea is to generate deeply optimized query plans tailored to one particular query and data stream. 
Since streaming queries are long-running~\cite{googleDataflow}, the high upfront effort for over-specialization is tolerable, and it differs significantly from traditional database optimizers, which must produce plans quickly~\cite{Marcus2023LearnedQS}. 
By spending more effort offline, \system{} can leverage time-consuming techniques for all optimization steps and synthesize bespoke query plans. 
This shift from fast plan generation to deep plan synthesis opens an entirely new frontier in stream optimization.

\begin{figure*}[h]
  \centering
  \includegraphics[width=\linewidth]{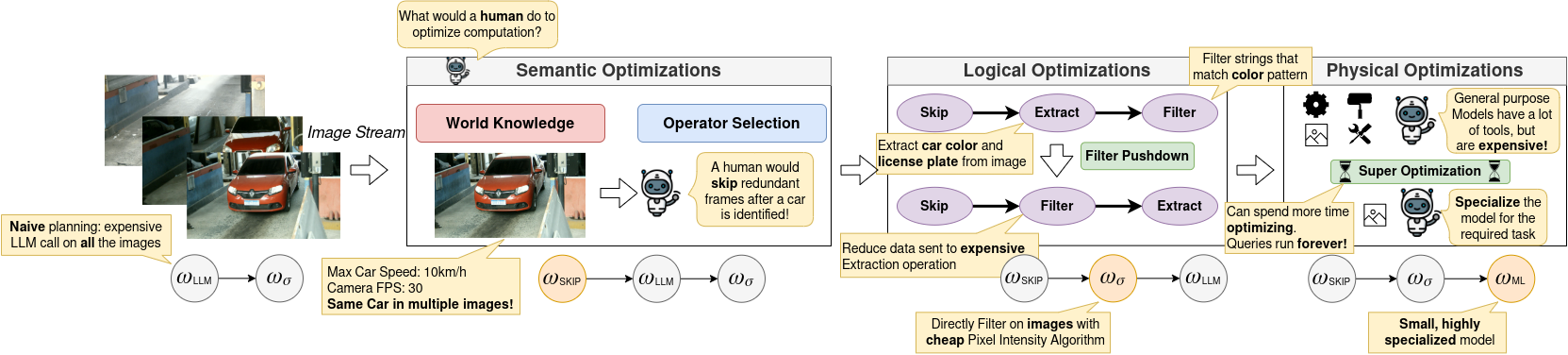}
  \vspace{-5ex}
  \caption{Overview of \system{}, a super-optimizer to enable multimodal streaming systems which transforms a naive multimodal query plan into an optimized plan through a series of novel semantic, logical, and physical optimizations which aggressively optimize the plan for a particular long-running streaming query plan.}
  \vspace{-3ex}
  \label{fig:samsara_overview}
\end{figure*}

\textbf{Anatomy of a Super-Optimizer.} 
We envision \system{} as a rethinking of query optimization for multimodal streaming shown in \Cref{fig:samsara_overview}. 
While some phases of optimization (logical and physical) are known from classical query optimization, a super-optimizer needs to rethink these phases and even add new phases.
One key innovation is a new class of \emph{semantic optimizations}, which leverage the world knowledge embedded to significantly reduce the volume of information processed by expensive MLLMs.
The main idea is that semantic optimization rewrites a plan in a manner similar to how a human expert with a deep understanding of the domain would approach it.
For example, by reasoning about car speed, camera frame rate, and prior vehicle positions as shown in \Cref{fig:samsara_overview} (left) for the traffic monitoring scenario, the optimizer can infer which upcoming frames will contain no new vehicle and skip them entirely---eliminating unnecessary inference. 

\textbf{The Challenge of Semantic Optimization.} 
While humans can manually identify such opportunities, our goal is to automate and generalize this process. 
We propose to leverage MLLMs themselves as reasoning agents within the optimizer: extracting world knowledge, inferring latent dependencies, and suggesting data-reduction transformations for arbitrary queries and datasets. 
This allows the optimizer to insert operators that prune redundant data and computation before they reach expensive AI operators. 
The central challenge lies in determining such semantic reductions automatically, without human input.
As we discuss later, the \system{} optimizer therefore follows a new optimization procedure where it first uses an LLM to understand the query and data, then it selects appropriate data reduction operators from a given catalog (e.g., frame skipping) to implement the derived optimization---essentially automating what a domain expert would design manually. Finally, it applies the selected operators iteratively in the plan. 

\textbf{Bespoke Logical \& Physical Optimizations.} 
Beyond semantic optimizations, \system{} applies novel logical and physical optimizations tailored to multimodal workloads as shown in \Cref{fig:samsara_overview} (right). 
For example, it can logically rewrite the query plan and push down the filter for only processing red cars.
The challenge here is that this filter must be cheaper to evaluate (i.e., without an MLLM), e.g., by using simple computer vision algorithms to detect that the image contains sufficient areas of red-ish pixels in our running example.
Another interesting direction is that we can afford to spend significantly more time on comprehensive optimizations. During physical optimization, this enables the integration of expensive techniques such as model pruning and distillation, which can drastically reduce the parameter size of MLLMs by tailoring them to data and one particular query. Although these techniques may take minutes or hours to apply, they can be executed offline before deploying the streaming query.
These combined optimizations allow \system{} to execute complex multimodal queries efficiently and at scale.

\vspace{-1ex}
\section{\system: A First Super-Optimizer}
\label{sec:case_study}

This section presents our super-optimizer prototype: \system. We show its design and functionality through a detailed case study that serves as a running example throughout this section. %While the explanations focus on this example, \system is not limited to a single query or dataset—its techniques generalize across workloads, as demonstrated in our initial evaluation later in this section.

\subsection{Case Study: Toll Booth} \label{subsec:case_study}

We demonstrate how semantic, logical, and physical optimizations can be applied in practice using a toll booth scenario with a live camera stream. Our initial evaluation further extends to a broader set of queries and datasets. To construct the case study, we created a dataset inspired by the Linear Road Benchmark, a well-established benchmark for streaming traffic monitoring data~\cite{arasu2004linear}. Specifically, we incorporated images from the Rodosol-ALPR dataset~\cite{laroca2022cross} to generate a video stream simulating cars passing through a toll booth, with each vehicle annotated by its license plate, color, and brand.
This scenario supports a variety of multimodal queries, such as filtering by specific vehicle attributes or counting vehicles to detect traffic patterns, and shows how \system optimizes different query types. For the purpose of the case study, we now focus on a specific query to demonstrate how it can be optimized using the principles of a super-optimizer. Consider the following scenario: \emph{A car has been reported stolen, and the police want to monitor all toll booths, each equipped with cameras. The available information indicates that the car is red and its license plate begins with “MTT.”} Based on this information, we can construct the naive multimodal query plan shown in \Cref{fig:a}, which first extracts the license plate and color, and then applies filters based on the given criteria.

\vspace{-2ex}
\subsection{Super-Optimizer Design}

In the following sections, we discuss using this example the design of \system\---our first super-optimizer. 

\begin{figure*}[h]% h=here, t=top, b=bottom, p=page
  \centering
  \includegraphics[width=\linewidth]{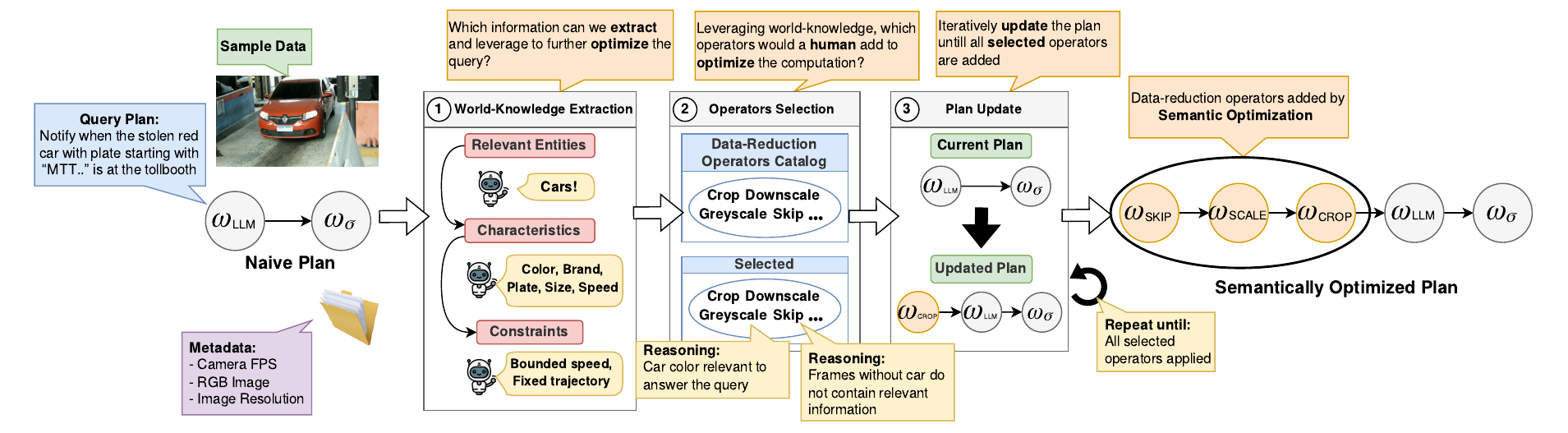}
  \vspace*{-5ex}
  \caption{Overview of the semantic optimization procedure applied to the tollbooth query. 
Starting from a naive plan (\emph{Extract + Filter}), the optimizer leverages an LLM to perform three reasoning phases: 
(1) \emph{World-Knowledge Extraction}, where it identifies relevant entities (cars), their properties (color, size, speed), and real-world dynamics (limited movement through the tollbooth); 
(2) \emph{Operator Selection}, where it proposes semantically motivated operators such as \texttt{Crop}, \texttt{Downscale}, and \texttt{Skip} based on extracted knowledge and metadata; and 
(3) \emph{Plan Update}, where the chosen operators are iteratively inserted to yield a semantically optimized plan (\emph{Crop + Extract + Filter}). 
The figure shows how LLM-guided reasoning bridges data semantics and query intent to automatically produce more efficient execution plans.}
  \label{fig:semantic-optimizer}
   \vspace{-3ex}
\end{figure*}

\vspace{-1.5ex}\subsubsection{Semantic Query Optimization}

Our vision introduces a new optimization phase, termed \emph{semantic query optimizations}, which uses world knowledge and contextual reasoning to rewrite multimodal streaming query plans in a human-like yet systematic fashion.  
Large language models (LLMs) play a central role in our framework. We leverage them throughout semantic optimization---to extract semantic priors, and to propose valid plan rewrites. This differs fundamentally from existing multimodal query systems, such as \textsc{CAESURA}~\cite{urban2023caesura}, which primarily employ LLMs to translate natural-language queries into executable plans. In contrast, we harness the reasoning capabilities of LLMs \emph{after} a plan has already been constructed, to optimize it semantically rather than syntactically.  

The key challenge is automation: how can such reasoning be systematically applied to arbitrary queries and data streams?  
We propose a semantic optimization procedure that takes as input (i) a data stream sample and (ii) the query plan; then, it uses an LLM-guided reasoning loop to identify optimizations that reduce redundant processing while preserving correctness.  
Figure~\ref{fig:semantic-optimizer} shows the stages of this process, from world-knowledge extraction to operator selection and plan rewriting, applied to the tollbooth case.

\textbf{Overview of the Optimization Procedure.}
Naively prompting an LLM to ``improve'' a query plan typically fails, since it lacks structured context and domain constraints.  
Instead, our optimizer uses the LLM as a semantic reasoning engine embedded in a structured three-phase process:  
\emph{(1) world-knowledge extraction},  
\emph{(2) operator selection}, and  
\emph{(3) plan update}.  
Each phase invokes targeted LLM prompts with structured inputs---a short query description, the plan operators, and sampled data summaries---so that the model’s reasoning remains grounded and verifiable.

\emph{(1) World-Knowledge Extraction.}
Given a query and a representative data sample, the optimizer first invokes the LLM to identify relevant entities, relationships, and constraints implied by both.  
For the tollbooth example, the query is:  
\emph{``Notify when the stolen red car with plate
starting with “MTT” is at the tollbooth''}  
The sample consists of short video segments from a fixed camera observing the scene.  
From this, the LLM extracts domain-specific priors, such as the fact that cars move approximately in a straight line through the tollbooth at bounded speeds, that empty frames frequently occur between cars, and that license plates and car colors are confined to specific spatial regions of the image.  
These extracted semantics extend the optimizer’s reasoning context, forming a symbolic representation of the scene guiding the subsequent steps. %\todo[inline]{This is really cool, and I think it should reflect (at least terminological) in Figure 3.}

\emph{(2) Operator Selection.}
Using the extracted symbolic representation of the scene, the optimizer next invokes the LLM to reason about which rewrites can safely reduce input volume or operator cost without affecting query correctness.  
In \system, we currently implement this as a selection procedure of data reduction tools from a given catalog of tools. Selecting the tools involves both \emph{cross-frame} and \emph{intra-frame} reasoning.  
In the cross-frame dimension, the LLM infers the temporal continuity of the scene; thus, cars cannot appear or disappear instantaneously.  
It therefore proposes a \texttt{Skip(Amount, Condition)} operator that skips $N$ frames after an empty detection, estimating $N$ from metadata such as frame rate and maximum vehicle velocity.  
For example, with a frame rate of 30 FPS and $v_{max}=30$ km/h, skipping more than three consecutive empty frames risks missing a new fast-approaching car.  
In the intra-frame dimension, the LLM infers that cars predominantly appear in the lower region of the frame and that color is the only query-relevant feature.  
It thus proposes a \texttt{Crop(region=bottom)} operator to restrict processing spatially, and a \texttt{Downscale(resolution)} operator to reduce pixel density while preserving color fidelity.  
However, the LLM explicitly rejects \texttt{Greyscale()} reduction, correctly reasoning that it would remove color cues critical to the query semantics.  
At this stage, the optimizer holds a set of validated candidate operators, each annotated with semantic preconditions. % and expected performance benefits.

\emph{(3) Plan Update.}
Finally, the optimizer integrates these operators into the query plan.  
Here again, the LLM assists by reasoning about operator dependencies and insertion points.  
For the tollbooth query, it proposes to insert \texttt{Skip(Amount=3, Condition=no\_car)} before the object detector to prune empty frames, and adds the operator \texttt{Crop(region=bottom)} before detection to restrict the spatial focus and to reduce computational overhead.  
The resulting plan, shown in Figure~\ref{fig:semantic-optimizer}, demonstrates the transformation from a naive syntactic plan into a semantically optimized one that minimizes redundant processing while maintaining correctness.  

\textbf{Generalization to Other Queries.}
While we have explained the procedure for the case study query, the same reasoning loop generalizes across domains, and in fact, we use this loop for all 13 queries over 2 different data streams in our initial evaluation. 
Consider a sports analytics query detecting which player currently possesses the ball in a soccer match.  
The LLM infers that ball possession transitions cannot occur instantaneously unless another player is nearby, allowing several frames to be skipped after a stable possession detection.  
Using the same extract–select–update framework, the optimizer thus introduces \texttt{Skip} and \texttt{Crop} operators driven by semantic understanding rather than syntactic structure.  

\textbf{Correctness of Rewrites.}
A core technical challenge lies in verifying the correctness of semantic rewrites.  
Determining whether a skip factor or downscaling level preserves query equivalence requires reasoning about both statistical fidelity and logical semantics.  
We employ an \emph{empirical validation} step in which the optimizer executes both the naive and the optimized plans on a data sample and compares their outputs to estimate accuracy degradation.  
This feedback loop transforms the optimizer into a self-correcting agent---capable of hypothesizing, testing, and refining its own rewrites through LLM-guided reasoning.  
In contrast to systems like \textsc{CAESURA}~\cite{urban2023caesura} that use LLMs to construct executable plans from natural language, our approach establishes a new paradigm of \emph{semantic optimization}, where LLMs reason about meaning to improve efficiency without altering intent.

\begin{figure*}[t]
    \centering
    \includegraphics[width=0.62\linewidth]{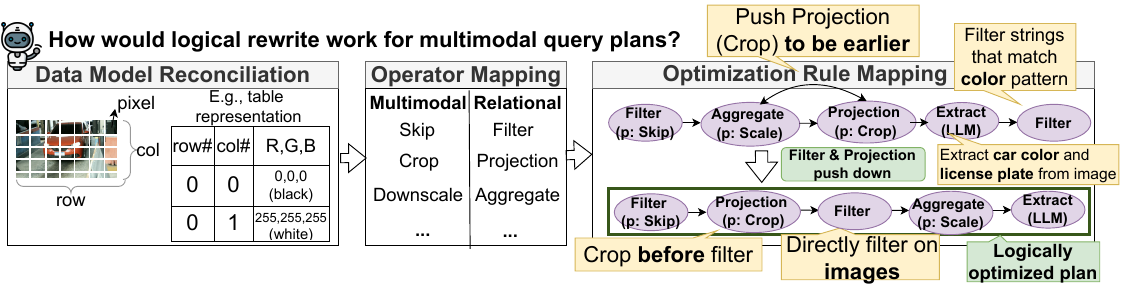}
    \vspace*{-3ex}
    \caption{Overview of logical optimiseation in \system. It starts with a semantically optimized plan with \texttt{Skip, Scale, Crop, MLLM, and Filter} operators to optimize it analogously to the relational counterparts, i.e., (1) Data model reconciliation that maps images to tuples, (2) Operator mapping that maps multimodal operators to relational and (3) Optimization rule mapping that applies logical rewrite rules. The resulting logically optimiseed plan is highlighted on the right.}
    \label{fig:logical_optimizer}
    % \vspace*{-3ex}
\end{figure*}

\vspace{-1.5ex}\subsubsection{Logical Optimization}\label{subsec:logical-opt}

Logical optimization in traditional databases rewrites query plans using heuristics such as filter pushdown or projection pushdown to improve efficiency while preserving correctness. In \system, we aim to achieve a similar effect for multimodal streaming plans, where operators act on images, text, or audio instead of structured tuples.

The challenge is that standard rewrite rules do not exist for multimodal operators. To address this, we leverage known relational optimization techniques and map them to multimodal data. For example, in our running scenario, a filter selecting frames with red cars---originally applied after an expensive MLLM-based \textit{Extract} operator---can be pushed earlier in the plan. Unlike structured data, this requires transforming the filter from a predicate on extracted attributes into a low-cost operation on the raw images. If this transformation is too expensive, the early filter could increase overall cost, which naturally connects logical optimization with physical optimization, as algorithm selection impacts efficiency.

\system leverages LLMs to reason about low-overhead operators' instantiation. For instance, the red-car filter allows for either a similarity search over embeddings or a simpler computer vision classifier. Other relational rules, such as projection or aggregation pushdown, can also be adapted: projection pushdown can correspond to cropping only the relevant parts of an image needed for downstream extraction, such as the license plate and car color.

The logical optimization is threefold as shown in \Cref{fig:logical_optimizer}. First, \emph{Data Model Reconciliation} provides a relational interpretation of the data modality. For images, pixels are mapped to tuples with schema $(\text{row\_id}, \text{column\_id}, \text{red}, \text{green}, \text{blue})$, with row and column forming a composite primary key. Second, \emph{Operation Mapping} translates multimodal operators to relational analogs: \textit{Crop} aligns with projection, while \textit{Downscale} corresponds to aggregation. Third, \emph{Optimization Rule Mapping} applies relational rewrite rules: for example, the \textit{Crop} operator can be pushed before \textit{Downscale}, mirroring filter pushdown. Similarly, selection predicates on multiple attributes (e.g., color and license plate) can be split, pushing the less expensive filter earlier to reduce the workload of downstream operators.

These optimizations have been implemented in \system, significantly reducing the workload of expensive plans by plan rewrites. Defining a systematic mapping from relational rewrites to multimodal operators---and specifying how operators after rewrite (e.g., filters and projections) can be efficiently realized on unstructured data---remains an open research challenge.

\begin{table}
\centering
\footnotesize
\vspace{-3ex}
\begin{tabular}{cp{2.8cm}p{3.5cm}}
\toprule
\textbf{Query ID} & \textbf{Description of Query} & \textbf{MLLM Tasks} \\ 
\midrule
Q1 & Car brand recognition & Object detection \\ 
Q2 & Car color recognition & Color recognition \\ 
Q3 & License plate detection & Object detection,
text extraction \\ 
Q4 & Most popular brand \& color &  Color recognition, object detection, aggregation \\ 
Q5 & Most popular brand & Color recognition, object detection, aggregation \\ 
Q6 & Most popular color & Color recognition, aggregation \\ 
Q7 & Repeated car detection & Object detection, text extraction, aggregation\\ 
Q8 & Red stolen 'MTT' car & Color recognition, object detection,
text extraction, filtering \\ 
Q9 & Unique license plates & Object detection, text extraction, aggregation \\ 
\midrule
Q10 & Amount of jumping players & Action recognition, aggregation \\ 
Q11 & Most offensive team & Action recognition, aggregation\\ 
Q12 & Notify when someone spikes & Action recognition \\ 
Q13 & 3 most common actions & Action recognition, aggregation \\ 
\bottomrule
\end{tabular}
\caption{Overview of Queries. Q1-Q9 are in the Toll Booth dataset and Q10-Q13 on the Volleyball dataset.}
\vspace{-18pt}
\label{exp:queries}
\end{table}

\vspace{-1.5ex}\subsubsection{Physical Optimization}\label{subsec:physical-opt}

In databases, physical optimization requires selecting the most efficient algorithm to execute a logical operator. This step is important for multimodal streaming, where operators often rely on expensive AI models. Instead of always invoking a general-purpose MLLM, we can select a bespoke model or synthesize one tailored to the query and data at hand, allowing for \emph{super-optimization} of the query execution.
Such a step also includes reducing the computation and memory requirements of MLLM models. Techniques such as quantization, pruning, knowledge distillation, and low-rank factorization are employed to compress models while preserving functionality. In our running example, text extraction leverages a quantised MLLM, while YOLO pre-filters frames containing cars. Quantization reduced the MLLM's weights and activations to 8-bit integers, halving model size and memory bandwidth, while YOLO removed irrelevant frames early in the pipeline, improving inference speed and overall FPS.

Streaming data, however, introduces new opportunities and challenges for known AI techniques such as pruning. Unlike classical AI tasks with highly diverse datasets, streaming data is often highly similar but continuously evolving. We can make use of this fact for physical optimization. For example, let us look at model pruning. Model pruning is a technique to reduce the size and computational cost of the model by removing unnecessary parameters. Highly similar parameters, too aggressive static pruning, may over-specialize and degrade performance when data characteristics change. For example, data from a traffic camera scenario may switch from low to high-density traffic: pruning aggressively in high-traffic periods could reduce accuracy, while in low-traffic periods it is safe. \system addresses this with new \emph{adaptive pruning} strategies that adjust parameters, such as pruning rates, based on streaming data characteristics dynamically. Revisiting classical model-reduction techniques and adapting them to the uniform but evolving nature of streaming data is an interesting future direction.

Another key aspect of physical optimization is selecting the appropriate algorithm. A large MLLM can sometimes be replaced by a smaller or distilled model, trading off computational cost against accuracy. Optimizers must consider accuracy constraints: similar to multimodal data systems such as LOTUS or Palimpzest~\cite{liu2025palimpzest}, which optimize plans for batch processing, where users to specify required relative accuracy (e.g., 90\% of the large MLLM), enabling the system to select the most efficient model that meets the accuracy threshold, we plan to apply similar ideas to stream processing. However, as data may continuously change, the model selection may need to be adapted over time as well. Building such an accuracy-guided adaptive model selection for multi-modal streaming systems is another open challenge for developing a super-optimizer for multimodal streaming systems.

\begin{figure}
    \centering
    \includegraphics[width=\linewidth]{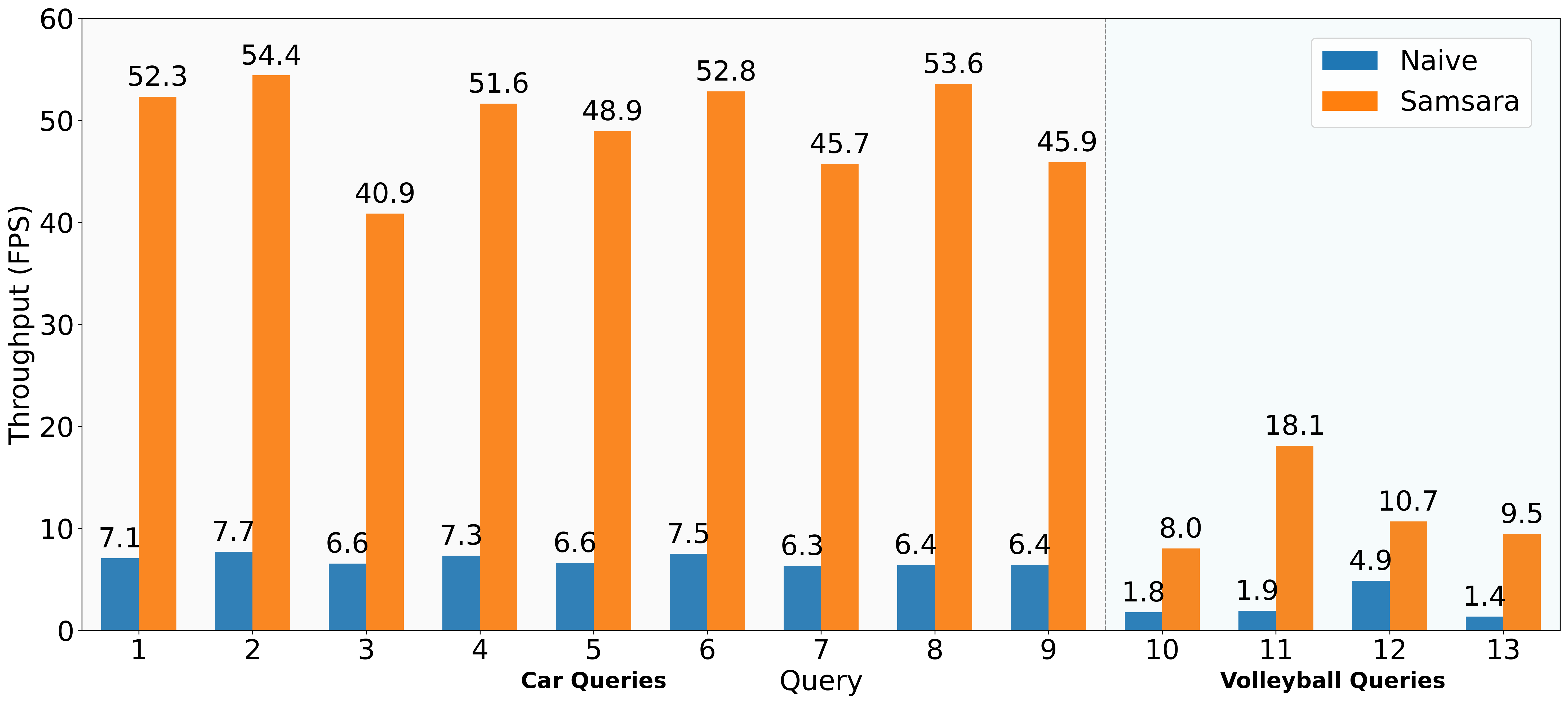}
    \vspace{-6ex}
    \caption{End-to-End gains for all(13) queries cf Table~\ref{exp:queries}. \system has up to $10\times$ speedup over naive execution}
    \vspace{-6ex}
    \label{fig:naive-vs-optimized}
\end{figure}

\subsection{Initial Results}

In the following, we show the initial results of using \system for multimodal stream processing. For this, we integrated \system with Apache Flink and executed queries without any optimizations (naive) and after optimizing plans with \system. 

\textbf{Dataset and Queries.}  
To the best of our knowledge, no benchmark currently exists for evaluating multimodal streaming queries. We therefore constructed a benchmark from existing datasets, focusing on two distinct domains that differ in complexity and data dynamics.
The first dataset, \emph{Toll Booth}, represents a static camera scenario where frames are captured from a fixed position and objects (cars) appear in predictable spatial regions. The second dataset, \emph{Volleyball}, is derived from the Volleyball Dataset~\cite{ibrahim2016hierarchical}, featuring moving cameras and multiple interacting and moving objects. 
Across both datasets, we defined thirteen queries as shown \Cref{exp:queries}: nine for the Toll Booth and four for Volleyball. Each query includes at least one multimodal LLM-based operator, targeting different extraction functions---such as \emph{color recognition}, \emph{object detection}, and \emph{text extraction}. In addition, the queries perform streaming-level operations such as \emph{aggregation} (e.g., counting objects over time windows) and \emph{filtering} over attributes (e.g., detecting cars of a specific color). The benchmark does not include joins between streams, which we plan to explore in future work. 

\textbf{Experiment 1: End-to-End Gains.}  
We evaluated \system across all thirteen queries, comparing the naïve query plan---where every frame is processed by a large multimodal LLM (Qwen2.5-VL)---against the optimized query plans that include semantic, logical, and physical optimizations. Throughput was measured in frames per second (FPS), while accuracy was computed at the query-result level.
Across all queries, \system achieved significant end-to-end speedups while the queries Q1-Q9 on the Tool Both data, which are less complex to optimize, show higher benefits than the Volleyball queries (Q10-Q13). However, all queries show significant throughput improvements, with the best case reaching up to \emph{$10
\times$ higher FPS}. On average, optimizations allowed several queries to reach or surpass real-time performance, turning previously infeasible pipelines into practical streaming deployments. Accuracy losses caused by model specialization and operator reordering were minimal, with a mean accuracy drop of \emph{7\%} relative to the baseline (naive execution). These results demonstrate that \system can substantially improve performance without sacrificing correctness.

\textbf{Experiment 2: Ablation Study.}  
To quantify the contribution of each optimization phase, we performed an ablation study summarizing the \emph{minimum}, \emph{average}, and \emph{maximum} FPS improvements across all thirteen queries. The results in Table \ref{tab:performance} confirm that each optimization phase---\emph{semantic}, \emph{logical}, and \emph{physical}---makes a meaningful contribution to overall performance, depending on the query and data characteristics.
Semantic optimizations produced the highest average and maximum gains by enabling early data reduction. Logical optimizations were most beneficial in cases where \emph{filter pushdown} could be applied and efficiently realized via low-cost operators. Physical optimizations further improved performance by selecting or adapting a lightweight model (i.e., the \emph{YOLOv8} object detector instead of the full-scale multimodal LLM (Qwen2.5-VL)).

\begin{table}
\centering
\small
\begin{tabular}{lccc}
\toprule
\textbf{} & \textbf{Min} & \textbf{Avg} & \textbf{Max} \\
\midrule
\textbf{Semantic}   & $1.9\times$ & $4.8\times$ & $8.0\times$ \\
\textbf{$+$Logical} & $2.1\times$ & $7.3\times$ & $10.1\times$ \\
\textbf{$+$Physical} & $2.3\times$ & $7.4\times$ & $10.4\times$ \\
\bottomrule
\end{tabular}
\caption{Speedup by optimization phases of \system{} over naive execution. We observe that across all queries, the minimum speedup is at least $2\times$. On average, we see around $6\times$ speedup, and at a maximum, $10\times$ speedup. We also observe that all phases are crucial for achieving these speedups, and that semantic optimization is particularly important. 
}
\vspace{-7ex}
\label{tab:performance}
\end{table}

\vspace{-3ex}
\section{Road Ahead}\label{sec:road_ahead}

This section outlines our research plan to evolve \system from a proof-of-concept to a principled, high-performance super optimizer for multimodal streaming systems.%, bridging formal foundations with deployable systems.

\noindent\textbf{Super-Optimization of Multimodal Streaming.}
While we have presented a first prototype of \system and demonstrated its promise, many research questions remain open. 
The next step is to transform semantic optimization---leveraging common-sense knowledge captured by LLMs---into a theory-backed, system-enforced capability that enhances continuous query plans without compromising declarativity. 
Key challenges include developing formal underpinnings for semantic rewrites with uncertainty-aware semantics, adapting classic planning problems such as cardinality estimation and plan enumeration to multimodal settings, and extending operator synthesis with specification-by-contract and cost/latency-aware search. 
An additional promising direction is \emph{adaptive model specialization}, where long-running queries with stable logic but evolving data (e.g., a fixed camera feed) allow lightweight retraining or pruning to yield faster and smaller models tailored to given workloads. All these directions aim to evolve \system into an autonomous, correctness-aware planner for multimodal streams.

\noindent\textbf{Multimodal Streaming Systems.}
While super-optimization is a key aspect of enabling the use of MLLMs for rich and efficient querying, integrating MLLMs into streaming systems opens many orthogonal directions for further exploration. 
Beyond optimization, other important aspects include \emph{semantic caching} to reduce redundant MLLM calls or even use caches for similar queries.
Moreover, we have integrated MLLMs as user-defined map operators into the query execution, but there are other opportunities, such as integrating MLLMs into more \emph{streaming operators} for multimodal data like joins over image streams to fuse streams across sources (e.g., merge two camera streams). 
Finally, 
% our work so far has focused primarily on image streams. As such,
extending these ideas to other modalities such as audio or other non-structured sensor data is an exciting next step. 
Although some techniques will transfer, new challenges will emerge across all layers---from optimization to operator design and beyond.

\noindent\textbf{Language Extensions and Novel Benchmarks.}
Integrating multimodality at the query language level is essential for declarative and efficient multimodal analytics. 
Future research will also focus on \emph{language extensions} that enrich continuous query semantics with multimodal predicates, uncertainty-aware constructs, and semantic annotations, forming the foundation for the super-optimization stack above. 
Finally, advancing this field requires robust \emph{benchmarks} and evaluation frameworks that provide annotated multimodal datasets and measure both throughput and semantic correctness, ensuring reproducible and comparable progress.

%\begin{acks}
%To Robert, for the bagels and explaining CMYK and color spaces.
%\end{acks}

\section*{Artifacts}
The source code and data used in this paper are available at: \\ 
\url{https://github.com/DataManagementLab/Samsara}

\bibliographystyle{ACM-Reference-Format}
\bibliography{software}

%\appendix
\end{document}